# Nine new M dwarf planet candidates from *TESS* including five gas giants

Yoshi Nike Emilia Eschen[1,2]★ and Michelle Kunimoto[1]†

[1]*Department of Physics and Kavli Institute for Astrophysics and Space Research, Massachusetts Institute of Technology, 77 Massachusetts Avenue, Cambridge, MA 02139, USA*
[2]*Department of Physics, University of Warwick, Gibbet Hill Road, Coventry, CV4 7AL, UK*

**ABSTRACT**
We present the detection of nine new planet candidates orbiting M dwarfs, identified using an independent search and vetting pipeline applied to *TESS* Full-Frame Image data from Sectors 1 to 63. Our candidates include planets as small as $1.4\,R_\oplus$, with orbital periods up to 20 d. Among the nine new candidates, we identified five gas giants, which represent a rare and unexpected outcome of planet formation. Our findings add to the growing sample of giant planets around M dwarfs found by *TESS*. We discuss their follow-up potential for mass measurements through radial velocity observations and atmospheric characterization through transmission spectroscopy. We highlight TIC 12999193.01 as a particularly unique gas giant candidate in an eccentric orbit and excellent potential for atmospheric characterization.

## 1 INTRODUCTION

The discovery of exoplanets has revolutionized our understanding of the prevalence, diversity, and formation of planetary systems. Among the myriad of exoplanetary hosts, M dwarfs have emerged as prime targets for exploration. These low-mass stars constitute the most abundant stellar population in the Milky Way, comprising around 70 per cent of all stars (e.g. Henry et al. 2018). Due to being smaller and cooler, the habitable zones of M dwarfs (where liquid water could exist on a rocky planet's surface) occur at much smaller orbital distances compared to Sun-like stars. Planets are also easier to detect around M dwarfs via the transit method due to the relatively large planet-to-star radius ratios, resulting in deeper transit depths.

Recognizing the unique advantages offered by M dwarfs, recent exoplanet research endeavours have increasingly focused on these low-mass stellar objects. In particular, NASA's Transiting Exoplanet Survey Satellite (*TESS*; Ricker et al. 2015) has played a pivotal role by monitoring the brightness of stars across nearly the entire sky since its launch in 2018. Tens of thousands of M dwarfs have received at least one sector (27 d) of observations at a cadence of two minutes, as processed by NASA's Science Processing Operations Center pipeline (SPOC; Jenkins et al. 2016), while millions more have data available at longer cadence (between 30 min and 200 s) through Full Frame Image (FFI) observations. *TESS* has detected ∼450 M dwarf planet candidates, including two Earth-sized planets in the habitable zones of M dwarfs (TOI-700d and e; Gilbert et al. 2020). Furthermore, among the most surprising *TESS* M dwarf planet discoveries have been a growing sample of gas giant planets. These are a particularly intriguing and rare phenomena, being significantly less common than hot Jupiters around Sun-like stars (Bryant, Bayliss & Van Eylen 2023; Gan et al. 2023). Planet formation models struggle to reproduce the formation of gas giants, around low-mass stars, especially in close-in orbits, due to lack of formation time and materials in their protoplanetary discs (Burn et al. 2021).

Thanks to its frequent and expedited public data releases, *TESS* is also a fantastic source of data for community-led searches to contribute new planets. Independent planet detection and vetting pipelines include the *TESS* Faint Star Search (Kunimoto et al. 2022), DIAmante (Montalto et al. 2020), the Convolutional Neural Networks search for Transiting Planet Candidates (Olmschenk et al. 2021), and NEMESIS (Feliz et al. 2021). NEMESIS was explicitly dedicated to searching for M dwarf planets with *TESS*, for which it focused on single-sector light curves of 33 054 M dwarfs from Sectors 1–5.

In this work, we describe a new independent search for planets around over 100 000 M dwarfs with *TESS*, based on multisector FFI observations across Sectors 1–63. We build off vetting tests designed for the Faint Star Search and add new tests to distinguish planets from false positives. We present nine new planet candidates obtained from this pipeline and their analysis.

## 2 METHODS

### 2.1 Light-curve generation

We started with all 117 475 M dwarfs ($R_* < 0.6\,R_\odot$, $M_* < 0.6\,M_\odot$, $2400 < T_{\rm eff} < 3900$ K) brighter than $T = 13.5$ mag in the *TESS* Input Catalogue (TIC) Candidate Target List (CTL) v8.01 (Stassun et al. 2019). Of our initial M dwarfs, 107 303 had light curves from the Quick-Look Pipeline (QLP; Huang et al. 2020) between sectors 1 and 63. QLP produces light curves using an aperture photometry approach combined with difference imaging, and has extracted light curves for all stars brighter than $T = 13.5$ mag since the start of the *TESS* mission. The QLP data set is the largest single source of publicly available light curves from *TESS* Full Frame Images to date.

★ E-mail: yoshi.eschen@warwick.ac.uk
† Juan Carlos Torres Fellow

The light curves were merged into multisector light curves to use all available data. We removed data flagged as poor quality (data with non-zero quality flags), resulting in stars having 17 per cent of data removed on average. We then detrended the light curves using the biweight algorithm implemented in WOTAN (Hippke et al. 2019), using a detrending width of 0.75 d. Given that M dwarf planet transits are typically a few hours or shorter, this window was chosen to minimize the impact of transit distortion caused by detrending while removing long-term astrophysical trends.

## 2.2 Planet search

We searched each multisector light curve using the Box-Least Squares (BLS: Kovács, Zucker & Mazeh 2002) algorithm implemented in CUVARBASE (Hoffman 2022), which uses GPUs to speed up time series analysis. CUVARBASE also adopts a Keplerian assumption to further optimize transiting planet searches, where only transit durations near the duration expected for a central, circular orbit at a given period and host star density are searched. The minimum period searched was set to 0.5 d, which was chosen to avoid unphysical orbits and reduce false positive contamination from stellar variability and contact binary stars at very short periods. The maximum period searched was set to half of the length of the longest continuous stretch of data (without sector-long gaps) or 100 d, whichever was smaller. The median star in our sample was searched for periods up to 25 d. To identify possible transit events, we required at least three transits and a signal-to-pink noise ratio (Hartman & Bakos 2016) greater than 9. A total of 13 642 signals met these detection criteria.

## 2.3 Candidate vetting

We based our vetting procedure on the automated vetting described by the *TESS* Faint Star Search (Kunimoto et al. 2022), which has discovered more than 3000 new *TESS* Objects of Interest (TOIs) to date. In summary, the Faint Star Search vetting procedure automatically identifies astrophysical false positives (eclipsing binaries and nearby eclipsing signals) and false alarms (systematics, noise, and stellar variability) by searching for sinusoidal variations, similar events at the same time-scale and significance of the candidate, significant odd–even transit depth differences, secondary eclipses, and highly grazing/large eclipses. The vetting procedure also produces and analyses difference images to identify off-target eclipsing signals (Bryson et al. 2013) using the TRANSIT-DIFFIMAGE codebase.[1]

After applying these tests to the signals meeting our detection criteria, we were still left with 2928 passing signals (Fig. 1), the vast majority of which should not be transiting planets. However, the Faint Star vetting procedure was originally designed to run after the ASTRONET-TRIAGE machine learning classifier used by QLP (Yu et al. 2019; Tey et al. 2023) had removed most false alarms, and thus it is currently unsuitable to be a stand-alone vetting pipeline. We added new steps to further automate this procedure, which were run on all 13 642 signals alongside the Faint Star tests:

---
[1] https://github.com/stevepur/transit-diffImage

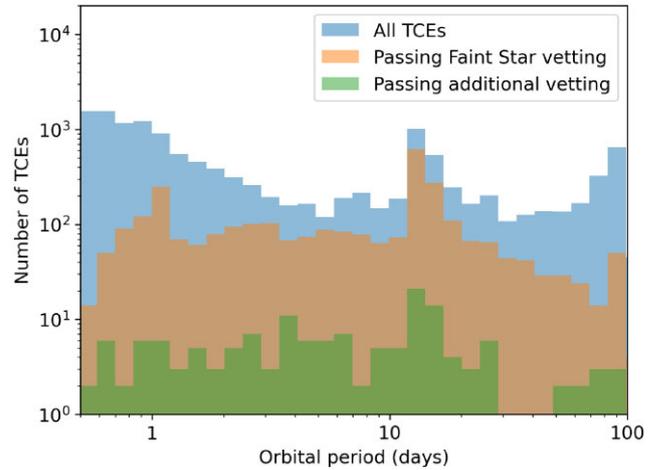

**Figure 1.** Period histogram of all TCEs passing basic detection criteria (blue). Roughly 20 per cent of TCEs passed the Faint Star vetting procedure described by Kunimoto et al. (2022) (orange), which was designed to remove astrophysical false positives following false alarm rejection by ASTRONET-TRIAGE (Tey et al. 2023). Only ∼1 per cent of all TCEs passed both the Faint Star vetting and our additional automated tests targeting false alarms (green). While we still see a systematic pileup near ∼13.7 d (the *TESS* orbit), our final passing TCEs are more uniformly spread across orbital period. These TCEs were then assessed through manual vetting (Section 2.4).

(i) We fit a trapezoid model to each signal, in addition to the transit models already fit in the Faint Star process.

(ii) We added new tests to remove signals with unphysical orbits, which are likely false alarms due to stellar variability.

(iii) We added a transit asymmetry test to identify false alarms that do not resemble symmetric, transit-like events.

These tests are described in more detail below.

### 2.3.1 Trapezoid model fit

The Faint Star procedure fits a transit model to each signal to derive important physical and orbital parameters, as well as to produce goodness-of-fit metrics to assess the consistency of the signal with a transit model. While we still adopted this step in the present work, we found that transit model fits would occasionally fail on signals that were clearly non-planet-like, had low signal-to-noise ratios, or were strongly affected by systematics. We found that trapezoid model fits were more robust to diverse signal shapes, providing us with useful metrics to further identify false positives.

The trapezoid model was parametrized by orbital period ($P$), transit epoch ($t_0$), transit depth ($\delta$), the transit duration divided by the orbital period ($q_{tran}$), and the transit ingress duration divided by the full transit duration ($q_{in}$). A central, box-shaped transit should have a $q_{in} \sim 0$, while a grazing, V-shaped transit should have $q_{in} \sim 0.5$. To speed up the fit process, data more than two transit durations from the BLS-inferred centre of each transit were ignored.

The results of the trapezoid model fit were used for new tests, as follows.

### 2.3.2 Unphysical orbit test

The transit model fit gives the ratio of the semimajor axis of the planet's orbit to the stellar radius, $a/R_s$. We removed 2401 signals where the transit model fit returned $a/R_s < 1.5$, since these signals

### 2.4 Manual triage

A total of 113 candidates survived the automated vetting procedure, 85 of which corresponded to known TOIs. Our pipeline did not flag any TOIs as false positives, with the exception of TOIs already dispositioned as false positives since being alerted.

We manually inspected the remaining 28 candidates by reviewing both flux and pixel-level diagnostics, and flagged another 17 signals as false positives. The majority of flagged signals were either borderline systematic or noise-like signals that were difficult for the vetter to distinguish from weak planet transit signals, or nearby eclipsing binaries (NEBs) in the same pixel as the target star which passed the automated check for centroid offsets. In summary, nine non-TOI M dwarf planet candidates remained after visual inspection, with difference images indicating that the candidates are consistent with being colocated with their target stars (Fig. 3).

### 2.5 Transit model fitting

We fit transit models with EXOPLANET (Foreman-Mackey et al. 2021) in order to obtain more accurate physical and orbital properties for the nine new candidates. We used initial guesses obtained from the transit model fits used in automated vetting, and ran four chains with 2000 tuning steps and 2000 draw steps each. We confirmed that all chains converged according to the Gelman–Rubin convergence statistic for each parameter satisfying $\hat{r} < 1.01$ (Gelman & Rubin 1992). The fitted properties are shown in Table 1, and final phase-folded light curves with the transit models overplotted are shown in Fig. 4. The planet radius and semimajor axis for each candidate was derived using the transit model fit results and stellar properties provided in the TIC catalogue (Table 2).

### 2.6 False positive probabilities

To increase confidence in each candidate, we ran the TRICERATOPS statistical validation package (Giacalone & Dressing 2020; Giacalone et al. 2021) to estimate each planet's False Positive Probability (FPP) and Nearby False Positive Probability (NFPP). Since we lack follow-up data for our candidates, we use TRICERATOPS not to claim that any of our candidates are statistically validated, but to further support the interpretation that these are candidates suitable for further follow-up and eventual confirmation.

Because our multisector light curves covered the Prime, First, and Second Extended Mission, which had different exposure times, the phase curves inputted to TRICERATOPS were binned to a 30 min cadence. Additionally, we only considered stars within 21 arcsec as possible NEB sources given that our difference images were able to constrain the source of each signal to within a *TESS* pixel. We computed FPPs and NFPPs 20 times, and report the mean of these values to obtain a result more robust to outliers, shown in Table 3.

We omit FPP results for our three candidates with $R > 8 R_{\rm Earth}$, as these are indistinguishable from brown dwarf and low-mass star scenarios based on *TESS* data alone, making their solutions degenerate and FPPs unreliable (Giacalone et al. 2021). All other candidates have low FPP (< 15 per cent) and very low NFPP (< 1 per cent), significantly improving confidence in their planet candidacy.

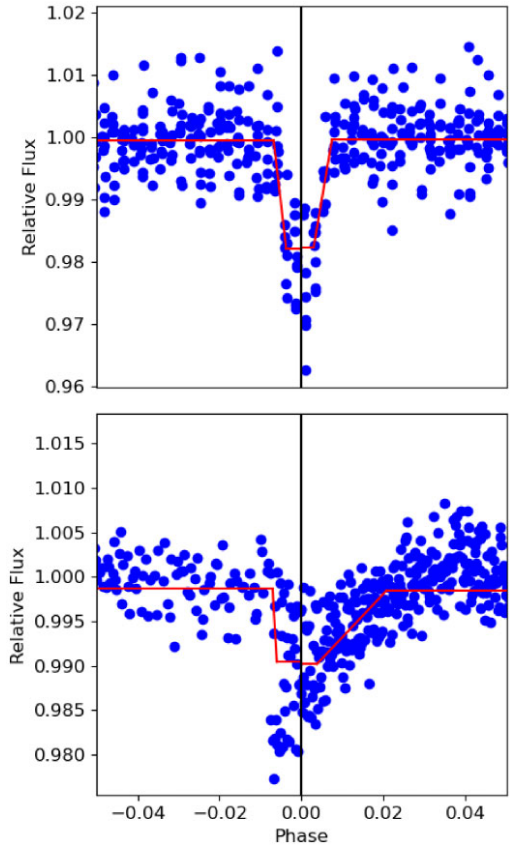

**Figure 2.** Top: an example of a symmetric transit, in this case the candidate around TIC 12999193. The red lines show half-trapezoid models separately fit to both left and right sides of the transit. Bottom: an example of an asymmetric signal caused by a systematic in the *TESS* data, with the associated half-trapezoid model fits.

would likely result in planets within the Roche limit from their stars and are hence in unphysical orbits. Although the minimum period searched by BLS should correspond to larger orbits for all stars in our sample, false positives can still return lower values through the transit model fit. Additionally, we found signals with durations significantly longer compared to our expectations for a circular orbit, which could indicate that the signal is caused by an eclipsing binary or stellar variability. We estimated the transit duration assuming a centrally transiting circular orbit and divided it by the orbital period to estimate $q_{\rm circ}$, and compared this to the $q_{\rm tran}$ value obtained from the trapezoid model fit. We removed 7760 signals with $q_{\rm circ}/q_{\rm tran} < 0.5$.

Finally, we flagged 141 signals with $q_{\rm tran} > 0.5$ as false positives, as this indicated that the signal occurred over half the orbital period, consistent with a contact binary system or stellar variability.

#### 2.3.3 Transit asymmetry test

A common feature of other false alarms was asymmetry in the transits, which we identified by fitting half-trapezoid models to the left and right sides of the transits, separately. For these fits, we set the period, transit time, and depth equal to the original trapezoid fit results and let $q_{\rm tran}$ and $q_{\rm in}$ vary. We labelled a signal as a false alarm if the significance of the difference between the left and right $q_{\rm tran}$ values was $> 5\sigma$, rejecting 764 signals. Examples of symmetric and asymmetric events are shown in Fig. 2.

## 3 RESULTS AND DISCUSSION

Our nine planet candidates have sizes between 2 and 11 Earth radii and orbital periods between 0.6 and 20 d Fig. 5. All candidates orbit host stars fainter than $T = 11$ mag, which is unsurprising given that brighter M dwarfs will have already been searched by the nominal QLP process ($T < 10.5$ mag). All host stars are in the cooldwarf catalogue (Muirhead et al. 2018), though follow-up spectroscopic observations would be valuable to better characterize the hosts and confirm their properties.

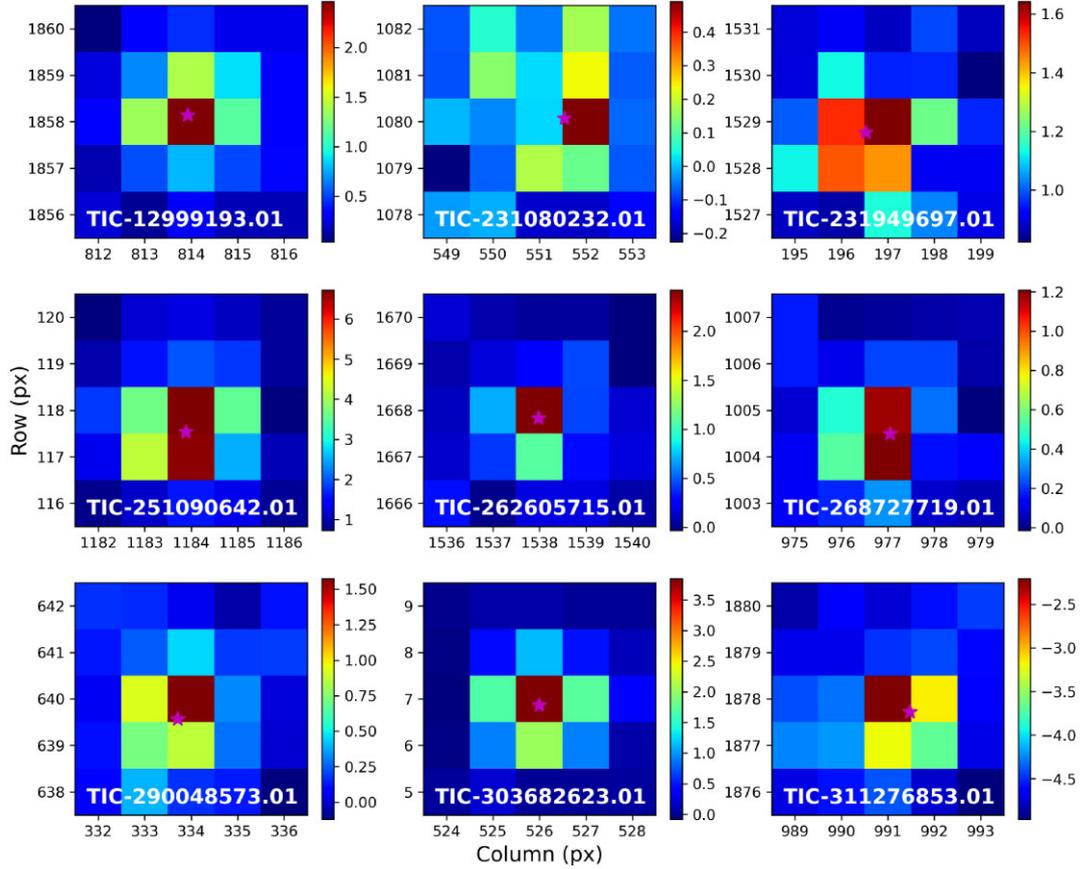

**Figure 3.** Difference images produced for each surviving candidate, made from the pixel time series of their most recently observed sectors. The pink star in each image indicates the location of the target star, showing that all passing candidates are consistent with on-target transiting planets.

**Table 1.** Fitted and derived parameters for the nine candidates based on exoplanet-fitted transit models.

| TIC ID | Period (d) | Epoch (BJD) | $R_p/R_s$ | Impact parameter | Duration (h) | Planet radius ($R_\oplus$) | Semimajor axis (AU) |
|---|---|---|---|---|---|---|---|
| 12 999 193 | $1.482641^{+0.000002}_{-0.000002}$ | $2458355.4444^{+0.0008}_{-0.0008}$ | $0.357^{+0.139}_{-0.095}$ | $1.146^{+0.159}_{-0.122}$ | $0.63^{+0.33}_{-0.23}$ | $9.78^{+3.83}_{-2.61}$ | $0.0154^{+0.0005}_{-0.0005}$ |
| 231 080 232 | $12.55434^{+0.00009}_{-0.00009}$ | $2458336.848^{+0.006}_{-0.005}$ | $0.037^{+0.002}_{-0.002}$ | $0.314^{+0.112}_{-0.172}$ | $2.80^{+0.14}_{-0.18}$ | $2.23^{+0.15}_{-0.15}$ | $0.086^{+0.001}_{-0.001}$ |
| 231 949 697 | $0.907204^{+0.000003}_{-0.000004}$ | $2458468.628^{+0.003}_{-0.003}$ | $0.026^{+0.002}_{-0.002}$ | $0.650^{+0.056}_{-0.103}$ | $0.88^{+0.06}_{-0.10}$ | $1.41^{+0.09}_{-0.11}$ | $0.0145^{+0.0002}_{-0.0002}$ |
| 251 090 642 | $20.04169^{+0.00004}_{-0.00004}$ | $2458879.409^{+0.001}_{-0.001}$ | $0.184^{+0.002}_{-0.002}$ | $0.063^{+0.064}_{-0.002}$ | $3.78^{+0.12}_{-0.12}$ | $10.42^{+0.34}_{-0.33}$ | $0.116^{+0.002}_{-0.002}$ |
| 262 605 715 | $1.161258^{+0.000002}_{-0.000002}$ | $2458517.845^{+0.0009}_{-0.0009}$ | $0.107^{+0.006}_{-0.005}$ | $0.819^{+0.031}_{-0.042}$ | $1.01^{+0.06}_{-0.07}$ | $6.85^{+0.461}_{-0.45}$ | $0.0180^{+0.0002}_{-0.0002}$ |
| 268 727 719 | $0.6256828^{+0.0000004}_{-0.0000004}$ | $2458766.5552^{+0.0006}_{-0.0006}$ | $0.112^{+0.007}_{-0.006}$ | $0.825^{+0.031}_{-0.042}$ | $0.84^{+0.05}_{-0.06}$ | $7.27^{+0.50}_{-0.43}$ | $0.0120^{+0.0001}_{-0.0001}$ |
| 290 048 573 | $1.483396^{+0.000004}_{-0.000003}$ | $2458326.656^{+0.001}_{-0.002}$ | $0.078^{+0.005}_{-0.005}$ | $0.767^{+0.045}_{-0.070}$ | $0.77^{+0.06}_{-0.08}$ | $2.86^{+0.21}_{-0.20}$ | $0.0174^{+0.0004}_{-0.0004}$ |
| 303 682 623 | $0.6806129^{+0.0000003}_{-0.0000003}$ | $2458438.5192^{+0.0005}_{-0.0004}$ | $0.153^{+0.008}_{-0.006}$ | $0.693^{+0.044}_{-0.037}$ | $1.03^{+0.05}_{-0.05}$ | $9.43^{+0.54}_{-0.49}$ | $0.0125^{+0.0002}_{-0.0002}$ |
| 311 276 853 | $5.16951^{+0.00003}_{-0.00003}$ | $2458987.476^{+0.003}_{-0.004}$ | $0.106^{+0.006}_{-0.007}$ | $0.435^{+0.109}_{-0.200}$ | $1.76^{+0.06}_{-0.06}$ | $4.71^{+0.29}_{-0.32}$ | $0.0430^{+0.0007}_{-0.0007}$ |

TIC 251090642.01, TIC 290048573.01, and TIC 303682623.01 are already known Community *TESS* Objects of Interest (CTOIs) listed on ExoFOP, and our derived parameters are within their reported errors. The other six candidates are new to this work.

### 3.1 New gas giant candidates

Five of our candidates may be gas giants based on their sizes ($R_p > 6$ $R_\oplus$; TIC 12999193.01, TIC 251090642.01, TIC 262605715.01, TIC 268727719.01, TIC 303682623.01). Gas giants around M dwarfs are relatively scarce compared to those around Sun-like stars, potentially suggesting that their formation involves different mechanisms or conditions than those around larger stars (Bryant et al. 2023; Gan et al. 2023). Current theories of planetary formation struggle to explain how gas giants could form in the relatively sparse protoplanetary discs that typically surround M dwarfs (Dawson &

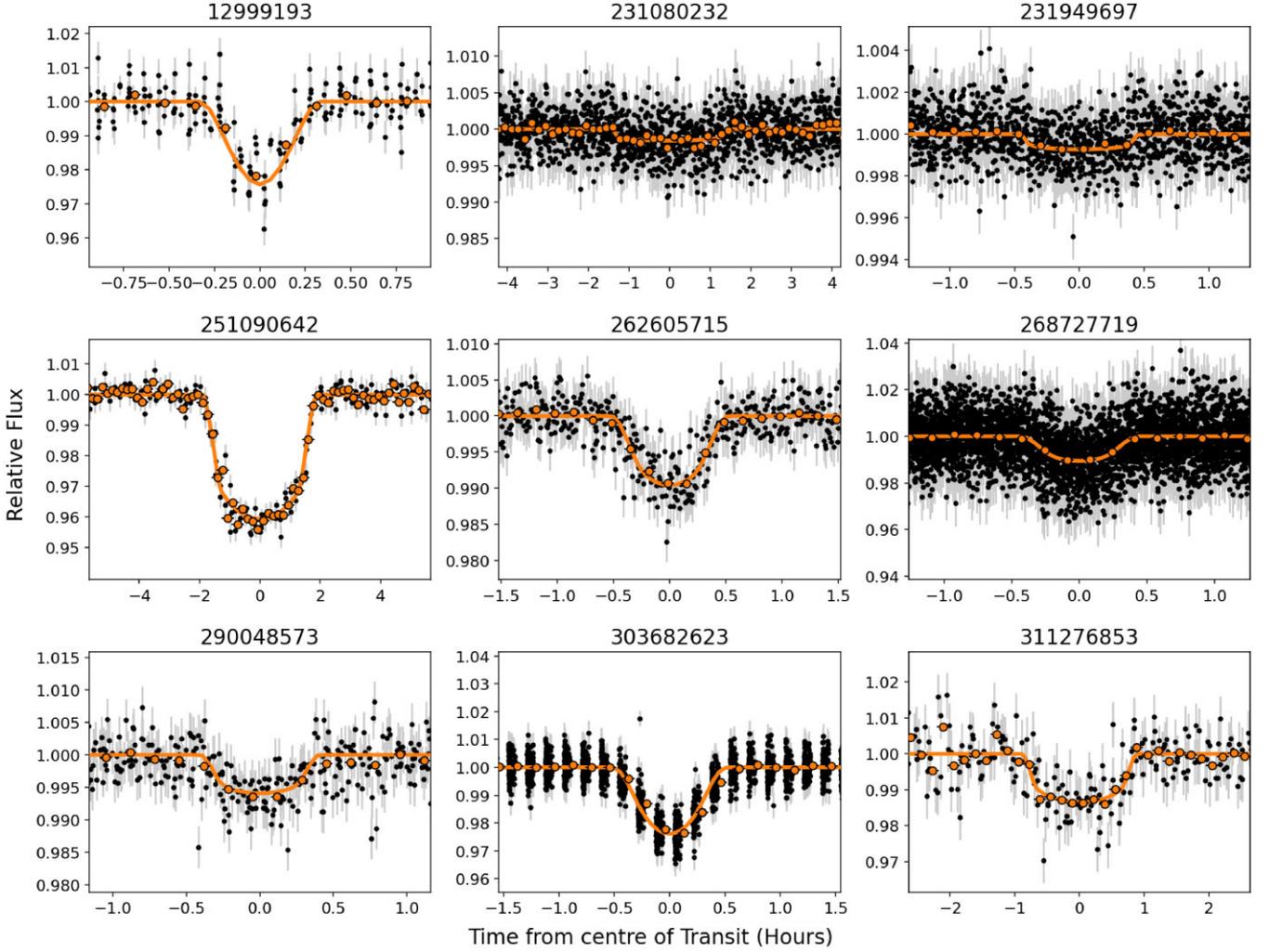

**Figure 4.** Phase curves for the nine planet candidates, showing *TESS* data in black. Orange lines show fitted transit models, while orange circles represent the underlying data averaged every 15 min.

**Table 2.** Stellar parameters obtained from the *TESS* Input Catalogue v8.01 (Stassun et al. 2019) for the host stars.

| TIC ID | RA (deg) | Dec (deg) | Tmag | Teff (Kelvin) | logg (logcm/s$^2$) | Mass (M$_\odot$) | Radius (R$_\odot$) |
|---|---|---|---|---|---|---|---|
| 12999193 | 346.857 | −28.354 | 13.366 ± 0.007 | 3478 ± 157 | 4.985 ± 0.013 | 0.222 ± 0.020 | 0.251 ± 0.008 |
| 231080232 | 68.018 | −68.883 | 12.870 ± 0.007 | 3812 ± 157 | 4.695 ± 0.010 | 0.544 ± 0.020 | 0.549 ± 0.016 |
| 231949697 | 94.258 | −43.366 | 11.330 ± 0.007 | 3701 ± 157 | 4.744 ± 0.008 | 0.491 ± 0.020 | 0.493 ± 0.015 |
| 251090642 | 139.794 | 53.595 | 13.091 ± 0.007 | 3560 ± 157 | 4.720 ± 0.009 | 0.518 ± 0.020 | 0.520 ± 0.015 |
| 262605715 | 132.45 | 0.207 | 12.555 ± 0.007 | 3618 ± 157 | 4.662 ± 0.011 | 0.577 ± 0.020 | 0.587 ± 0.018 |
| 268727719 | 347.957 | 65.23 | 13.389 ± 0.008 | 3718 ± 157 | 4.657 ± 0.011 | 0.582 ± 0.020 | 0.593 ± 0.018 |
| 290048573 | 319.189 | −25.786 | 12.680 ± 0.007 | 3518 ± 157 | 4.888 ± 0.001 | 0.318 ± 0.020 | 0.336 ± 0.010 |
| 303682623 | 75.769 | 14.356 | 13.018 ± 0.008 | 3469 ± 157 | 4.681 ± 0.010 | 0.558 ± 0.020 | 0.565 ± 0.017 |
| 311276853 | 247.739 | 26.107 | 13.240 ± 0.007 | 3583 ± 157 | 4.823 ± 0.004 | 0.396 ± 0.020 | 0.404 ± 0.012 |

Johnson 2018). These discs contain less material compared to those around more massive stars, making it challenging for gas giants to accrete enough mass (Burn et al. 2021).

Despite their rarity, *TESS* is finding a growing sample of giant planets around low-mass stars, especially hot Jupiters ($R_p > 8$ R$_\oplus$ and $P < 10$ d). To date, 11 M dwarf hot Jupiters have been confirmed, such as TOI-5205 b (Kanodia et al. 2023), HATS-71 (Bakos et al. 2020), and TOI-3235 (Hobson et al. 2023). If confirmed, our candidates would add another two hot Jupiters to this sample and provide new targets for further characterization. TIC-303682623.01 may also have the shortest orbital period of all known M dwarf gas giants to date, at $P = 0.68$ d.

A major challenge with identifying gas giants around M dwarfs is that such objects are more likely to be eclipsing binary false positives compared to those around Sun-like stars. Based on Gaia DR3 observations (Gaia Collaboration 2016, 2023), our giant candidate hosts TIC 12999193, 251090642, and 268 727 719 all have low Renormalized Unit Weight Error (RUWE) values (1.17, 1.09, 1.02

**Table 3.** TRICERATOPS results for FPP and NFPP of each transit signal. FPPs are only provided for planets smaller than $8\,R_\oplus$ due to degeneracies with brown dwarfs and low-mass stars.

| TIC ID | FPP | NFPP |
| --- | --- | --- |
| 12 999 193 | – | 0.0 |
| 231 080 232 | 0.031 | 0.0 |
| 231 949 697 | 0.122 | 0.002 |
| 251 090 642 | – | 0.0 |
| 262 605 715 | 0.0924 | 0.0 |
| 268 727 719 | 0.0364 | 0.0 |
| 290 048 573 | 0.030 | 0.006 |
| 303 682 623 | – | 0.0 |
| 311 276 853 | 0.036 | 0.0 |

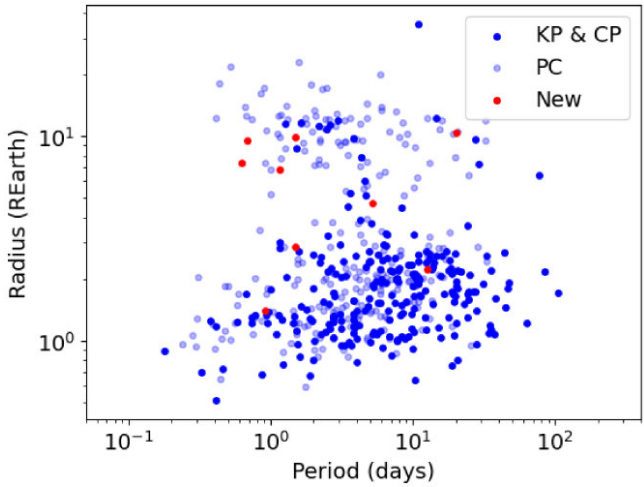

**Figure 5.** Distribution of planets and planet candidates around M dwarfs in radius-period space. Known and confirmed planets (KP & CP) were retrieved from the NASA Exoplanet Archive (NASA Exoplanet Archive 2024) (accessed on 2024 April 02) and plotted in blue, with planet candidates (PC) plotted in transparent blue. Our nine candidates are plotted in red.

consistent with being single stars (Lindegren et al. 2018, 2021). TIC 303 682 623 was missing a RUWE value, and TIC 262 605 715 has evidence for binarity given RUWE = 2.81 (> 1.4); nevertheless, we retain it as a candidate in case it is a planet in a two-star system given we cannot reject a planet scenario at the level of the data, and we re-emphasize that follow-up of all of our candidates will be needed for confirmation and better characterization of the host stars.

There are additional challenges associated with estimating accurate radii for objects grazing small hosts. Most of our giants have non-grazing impact parameters ($b < 0.9$), suggesting that their planetary radius estimates are robust. However, the candidate orbiting the smallest star among our new hosts (TIC-12999193.01) has a highly grazing impact parameter ($b = 1.1$), and its radius and duration estimates have large uncertainties, potentially further indicating an eccentric orbit. To explore the sensitivity of the transit model parameters to our assumption of circular orbits, we re-ran a further model fit using exoplanet, allowing eccentricity ($e$) and argument of pericenter ($\omega$) to vary. This fit converged at an eccentricity of $0.469^{+0.111}_{-0.093}$ and planet radius of $11.7^{+9.1}_{-5.4}\,R_\oplus$. We compared the eccentric and circular models based on the leave-one-out cross-validation information criterion (LOO; Vehtari, Gelman & Gabry 2015) as computed from the EXOPLANET results using the ARVIZ package for analysis of Bayesian models (Kumar et al. 2019).

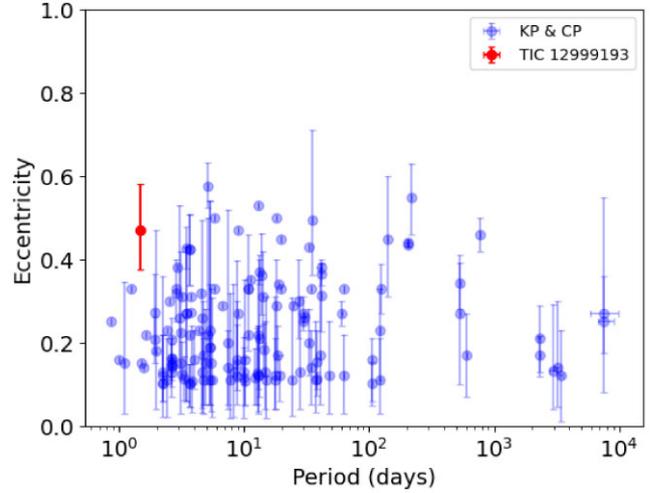

**Figure 6.** Eccentricities ($> 0.1$) of known and confirmed planets (KP & CP) around M dwarfs obtained from the NASA Exoplanet Archive (NASA Exoplanet Archive 2024; accessed on 2024 April 02) are plotted in blue. Our planet candidate is marked in red and has a period of 1.48 d.

LOO is a fully Bayesian model selection method that uses the entire posterior distribution. A model is favoured if it has a higher log(LOO) score. We found that the eccentric model was the preferred fit, with $\Delta \log(\text{LOO}) = 9.38 \pm 3.64$ over the circular model. This eccentricity would be relatively high compared to other M dwarfs planets near its orbit (Fig. 6). If TIC-12999193.01 is indeed a planet on an eccentric orbit, it will be an intriguing target for investigation into the high-eccentricity migration scenario for M dwarf hot Jupiter origins.

Despite the relatively high likelihood that TIC-12999193.01 is a stellar object due to its grazing orbit, we opted to keep it as a possible planet to be added among other known highly grazing and eccentric hot Jupiters, such as HIP 65 A b (Nielsen et al. 2020) and Kepler-447 b (Lillo-Box et al. 2015). Follow-up radial velocity data could confirm (or deny) its planetary nature and provide a significantly more accurate estimate of its orbital eccentricity.

More generally, follow-up observations could provide key clues to differentiate between the different formation mechanisms explaining these rare planets, beyond the goal of their confirmation. Measurements of their orbital eccentricities with radial velocity data could suggest histories involving high-eccentricity migration, while characterization of their atmospheres could suggest initial formation locations either close or far away from the host star. These planets will also be promising targets for investigations of star–planet interactions given their close-in orbits and much lower star–planet mass ratios compared to giants around Sun-like stars.

### 3.2 Follow-up potential

In order to check whether our candidates are suitable for radial velocity (RV) follow-up, we calculate their semi-amplitude $K$ using

$$K = 28.4 \left(\frac{M_\text{P}}{M_\text{J}}\right) \left(\frac{M_\text{P} + M_\text{S}}{M_\odot}\right)^{-1/2} \left(\frac{a}{1\text{AU}}\right)^{-1/2}, \quad (1)$$

where $K$ is in ms$^{-1}$ and the planet mass, $M_\text{P}$, is predicted via the mass–radius relations described by Chen & Kipping (2017). We also computed the Transmission and Emission Metric for each candidate (Kempton et al. 2018) to quantify their level of promise for atmospheric characterization through transmission and emission spectroscopy, respectively. The resulting RV semi-amplitudes, TSM, and ESM values are shown in Table 4.

**Table 4.** Computed semi-amplitude and Transmission and Emission Spectroscpy Metrics for each new planet candidate.

| TIC | Semi-amplitude (ms$^{-1}$) | TSM | ESM |
| --- | --- | --- | --- |
| 12999193 | 106.348 | 700.703 | 209.067 |
| 231080232 | 2.327 | 18.140 | 0.636 |
| 231949697 | 2.744 | 8.644 | 6.682 |
| 251090642 | 28.274 | 105.508 | 5.927 |
| 262605715 | 33.293 | 162.580 | 60.509 |
| 268727719 | 44.926 | 150.005 | 71.931 |
| 290048573 | 10.311 | 128.535 | 15.854 |
| 303682623 | 70.038 | 264.880 | 151.459 |
| 311276853 | 13.772 | 83.136 | 8.259 |

Our best follow-up target is the gas giant candidate TIC 12999193.01, which features a large, easily-detectable RV semi-amplitude ($K \sim 100$ ms$^{-1}$) and one of the ten highest TSM values of all known M dwarf planets and planet candidates (TSM = 701). The ultrashort period gas giant candidate TIC-303682623.01 is also a promising follow-up target ($K = 70$ ms$^{-1}$, TSM = 265), especially given that its impact parameter is well constrained and our estimate of its radius ($R_p = 9.4$ R$_\oplus$) is more reliable. Among smaller planets ($R < 4$R$_\oplus$), TIC-290048573.01 has $K = 10.3$ ms$^{-1}$ and TSM = 129, sufficient to be considered an excellent target for characterization of sub-Neptunes according to thresholds described by Kempton et al. (2018).

## 4 CONCLUSION

We present the detection of nine planet candidates found in a multisector search of 107 303 M dwarfs with *TESS* data from Sectors 1–63. 3 of these candidates are already known CTOIs, while six are new to this work. Among our planet candidates, we find five gas giants, adding to a growing sample of gas giants around M dwarfs found by *TESS* which are challenging theories of planet formation and are intriguing targets for follow-up. One of these candidates (TIC 12999193.01) also potentially has a high eccentricity ($e \sim 0.47$) and is hence an exciting target for further investigations into a possible origin through high-eccentricity migration. If confirmed, this candidate would be among the most eccentric short-period planets found around M dwarfs.

This work will be implemented into the Faint Star Search algorithm with further improvements to automate the entire vetting process, and to maximize planet yields from future community-led searches of *TESS* data.

## ACKNOWLEDGEMENTS

We thank the referee for their feedback which improved the manuscript and clarified text. This paper includes data collected by the *TESS* mission, which are publicly available from the Mikulski Archive for Space Telescopes (MAST). Funding for the *TESS* mission is provided by NASA's Science Mission directorate. We acknowledge the use of public *TESS* data from pipelines at the *TESS* Science Office and at the *TESS* Science Processing Operations Center. This research has made use of the Exoplanet Followup Observation Program website, which is operated by the California Institute of Technology, under contract with the National Aeronautics and Space Administration under the Exoplanet Exploration Program. This research has made use of the NASA Exoplanet Archive, which is operated by the California Institute of Technology, under contract with the National Aeronautics and Space Administration under the Exoplanet Exploration Program. MK acknowledges support by the Juan Carlos Torres post-doctoral fellowship from the MIT Kavli Institute for Astrophysics and Space Research.

## DATA AVAILABILITY

We make our fitted and derived planet parameters publicly available in a machine readable format. The light curves used for our search are already available as high-level science products (HLSPs) on MAST, from QLP.

## SUPPORTING INFORMATION

Supplementary data are available at *MNRAS* online.

**suppl_data**

Please note: Oxford University Press is not responsible for the content or functionality of any supporting materials supplied by the authors.

Any queries (other than missing material) should be directed to the corresponding author for the article.

This paper has been typeset from a T$_{\rm E}$X/L$^{\rm A}$T$_{\rm E}$X file prepared by the author.